\documentclass{rmf-d}
\usepackage{nopageno,rmfbib,multicol,times,epsf,amsmath,amssymb,cite}
\usepackage[latin1]{inputenc}
\usepackage[]{caption2}
\usepackage{float}
\usepackage{graphicx}
\usepackage{xspace}
\usepackage{lipsum}

\def\rmfcornisa{ \hfill\rmf\ {\bf } 
	\hfill }

\def\rmfcintilla{}
\clearpage \rmfcaptionstyle 
\begin{document}
\title{   Studying the mechanisms for strange particle production with ALICE at LHC  
\vspace{-6pt}}
\author{ Meenakshi Sharma, for the ALICE Collaboration      }
\address{ Department of Physics, University of Jammu    }
\address{meenakshi.sharma@cern.ch }
\author{ }
\address{ }
\author{ }
\address{ }
\author{ }
\address{ }
\author{ }
\address{ }
\maketitle
\recibido{}{
\vspace{-12pt}}
\begin{abstract}
\vspace{1em} \textbf{Abstract}: 
The main goal of the ALICE experiment is to study the physics of strongly interacting matter, focusing on the properties of the quark-gluon plasma (QGP). The relative production of strange hadrons with respect to non-strange hadrons in heavy-ion collisions was historically considered as one of the signatures of QGP formation. However, the latest results in proton-proton (pp) and proton-lead (p-Pb) collisions have revealed an increasing trend in the yield ratio of strange hadrons to pions with the charged-particle multiplicity in the event, showing a smooth evolution across different collision systems and energies. \\
We present the new studies which are performed with the aim of better understanding the production mechanisms for strange particles and hence the strangeness enhancement phenomenon in small collision systems. In one of the recent studies, the very forward energy transported by beam remnants (spectators) and detected by the Zero Degree Calorimeters (ZDC) is used to classify events. The contribution of the effective energy and the particle multiplicity on strangeness production is studied using a multi-differential approach in order to disentangle initial and final state effects. In the second study, the origin of strangeness enhancement with multiplicity in pp has been further investigated by separating the contribution of soft and hard processes, such as jets, to strange hadron production. Techniques involving full jet reconstruction or two-particle correlations have been exploited.
The results indicate that the increased relative strangeness production emerges from the growth of the underlying event, being disconnected from initial state properties, and suggest that soft (out-of-jets) processes are the dominant contribution to strange hadron production.

  \vspace{1em}
\end{abstract}
\keys{  strangeness enhancement, QGP, ALICE, Effective energy, multiplicity, LHC  \vspace{-4pt}}
\pacs{   \bf{\textit{25.75.-q, 25.75.Dw, 25,75.Gz, 25,75.Ld, 25,75.Nq     }}    \vspace{-4pt}}
\textbf{Introduction:}
\begin{multicols}{2}
\hspace{-16pt}Strangeness production is considered one of the important tools in the search and discovery of the primordial state of matter which existed a few microseconds after the Big Bang \cite{c1,ca1,ca2}. Rafleski and Muller \cite{c1} reported for the first time that the enhancement of the relative strangeness production could be one of the signatures of a phase transition from hadronic matter to the new phase consisting of almost free quarks and gluons (QGP). Strangeness enhancement was observed for the first time at the SPS \cite{c2}, then at RHIC \cite{c3} and later at the LHC \cite{c4} at increasing collision energies.
\begin{center}
  \includegraphics[scale=0.07685]{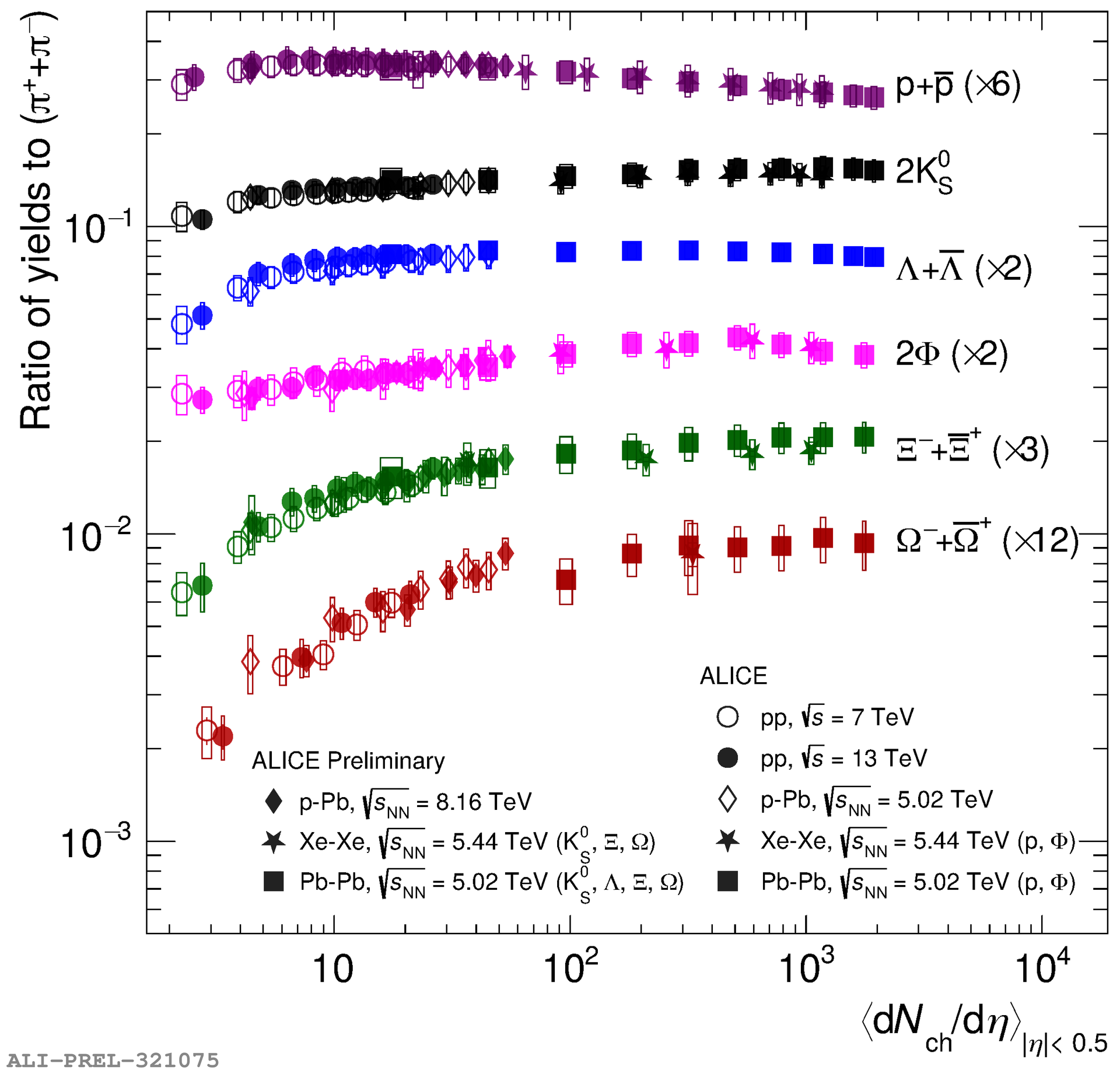}
	\flushleft \texttt{\small {FIGURE 1.} }\small {Ratio of hadron yields to pion yields in different collision systems at different collision energies}
\end{center}

\begin{center}
	\includegraphics[scale=0.125]{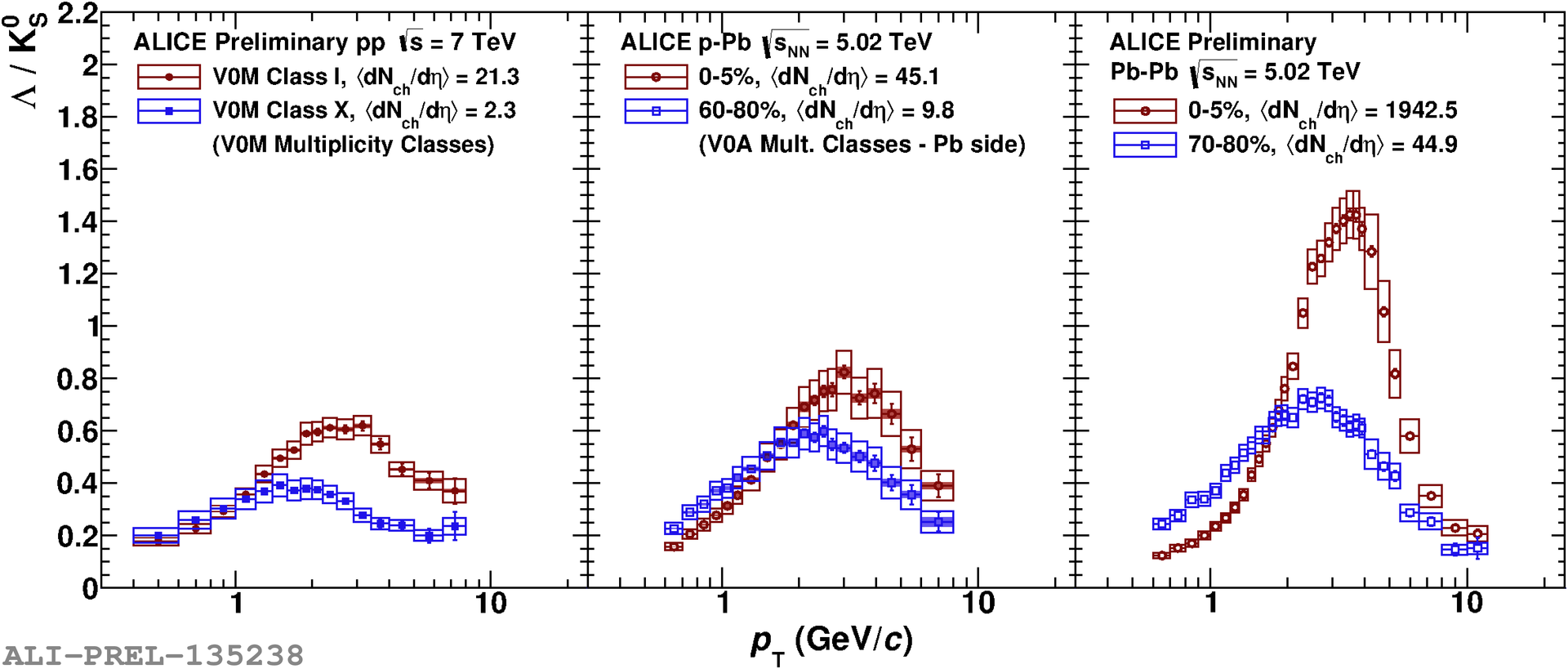}
	
\flushleft \texttt{\small {FIGURE 2.} }\small {Enhancement of baryon over meson ratio at intermediate $p_{\mathrm{T}}$ across different collision system }
\end{center}

\vspace{10pt}

\hspace{-16pt}The ALICE experiment has also studied strangeness production in different collision systems (pp, p-Pb and Pb-Pb) at different collision energies \cite{c7,c8,c9,c10}. The result in Fig. 1 shows that the ratio of strange to non-strange hadron yields increases with charged-particle multiplicity, showing a smooth evolution across different collision system and energies.
 It is also observed that the enhancement is larger for the particles with larger strangeness content. Several features that were observed in large collision systems and explained as due to the formation of the QGP or collective phenomenon are also observed in the small systems \cite{c12}. This includes the enhancement of $\Lambda/\mathrm{K^{0}_{s}}$ at intermediate $p_{\mathrm{T}}$ shown in Fig. 2 and the evolution of the particle spectra with multiplicity and hardening of the spectra towards higher multiplicity have also been observed. These unexpected results have motivated to perform studies using ALICE detector for understanding the mechanisms responsible for such behaviour in small systems.\\

	\textbf{Results:}\vspace{0.5cm}		
	
	\hspace{-16pt}In ALICE, strange hadrons $\mathrm{K^{0}_{s}}$, $\Lambda$, $\Xi$ and $\Omega$ are reconstructed using their weak decay daughter tracks in the central pseudorapidity region \cite{cm1}. The subdetectors involved in the analyses presented here include the Inner Tracking System (ITS) and the Time Projection Chamber (TPC) which are used for particle identification (PID) and tracking. Two forward dectectors placed on both sides of the ALICE interaction point, the VZERO and Zero Degree Calorimeter (ZDC) are used to classify events in multiplicity and energy event classes, respectively. A detailed description of the ALICE detector can be found in \cite{c5}. The two analyses which are presented here exploit the multi-differential approaches to explore strangeness enhancement in pp.\\
		\begin{center}
	\includegraphics[scale=0.11]{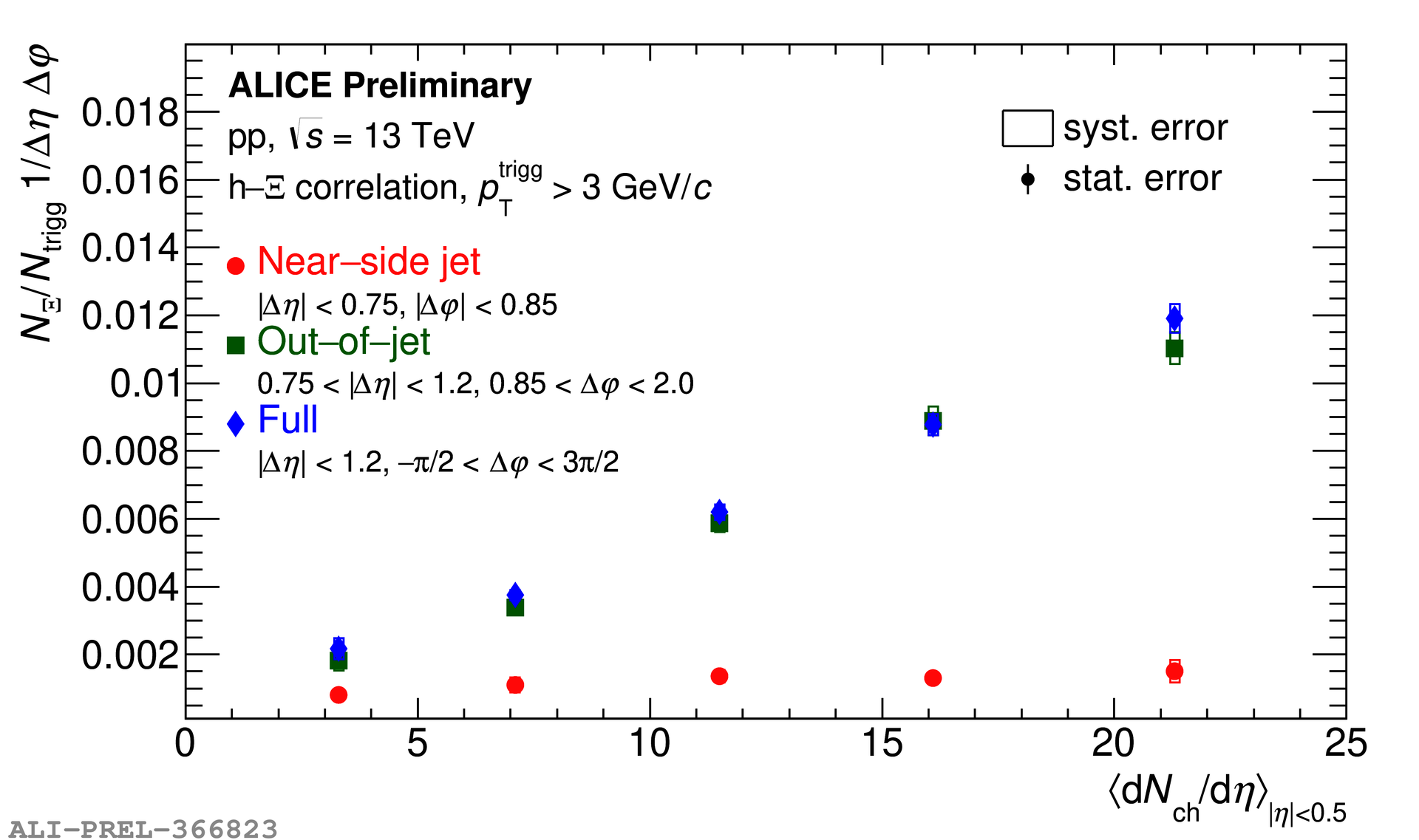}
	\flushleft \texttt{\small{FIGURE 3. }}\small{ $\mathrm \Xi$ yields per trigger particle and per unit of $\mathrm \Delta\eta\Delta\phi$ as a function of the charged-particle multiplicity produced at midrapidity. }
	\end{center}
\vspace{10pt}
	
\hspace{-16pt} The first analysis involves the angular correlation method to seperate $\mathrm{K^{0}_{s}}$ and $\Xi$ hadrons produced in jets (hard processes) from those produced out-of-jets (soft processes), where the hard processes are characterised by large momentum transfer. The particles produced in the near-side jet region are characterised by a small angular seperation from the leading particle of the jet, which is identified as the particle with the highest transverse momentum in the collision with $p_{\mathrm{T}}$ $>$  3 GeV/\textit{c}. The $\Xi$ yields per trigger particle and per unit of $\Delta\eta\Delta\phi$ are displayed in Fig. 3 as a function of the charged-particle multiplicity produced at midrapidity. The full and out-of-jet yields increase with multiplicity, while the near-side-jet yields show a very mild to no dependence on particle production at midrapidity. Similar results are also obtained for $\mathrm{K^{0}_{s}}$ yields. These results suggest that out-of-jet (soft) processes are the dominant contribution to strange particle production in pp collisions.\\
	
\hspace{-16pt}	The fraction of the initial energy spent in the hadronization process is known as effective energy. Due to the leading baryon effect, there is a high probablity to emit baryons in the forward direction with high longitudinal momentum, carrying away a considerable fraction of the total available energy. Therefore, the effective energy is always less than the initial collision energy \cite{c6}. In ALICE, the Zero Degree Calorimeters (ZDCs) are used to reconstruct the energy of the leading nucleons and define the effective energy classes using the relation, $\mathrm E_{eff} = \sqrt{s} - \mathrm E_{|\eta|>8}$. The events are classified in effective energy and multiplicity percentile classes, using the ZDC  and V0 detector, respectively. Both Monte-Carlo simulation and data have confirmed that the effective energy and multiplicity are correlated.\\
	
	\begin{center}
		\centering
		\label{fig:d}
		\includegraphics[scale=0.097]{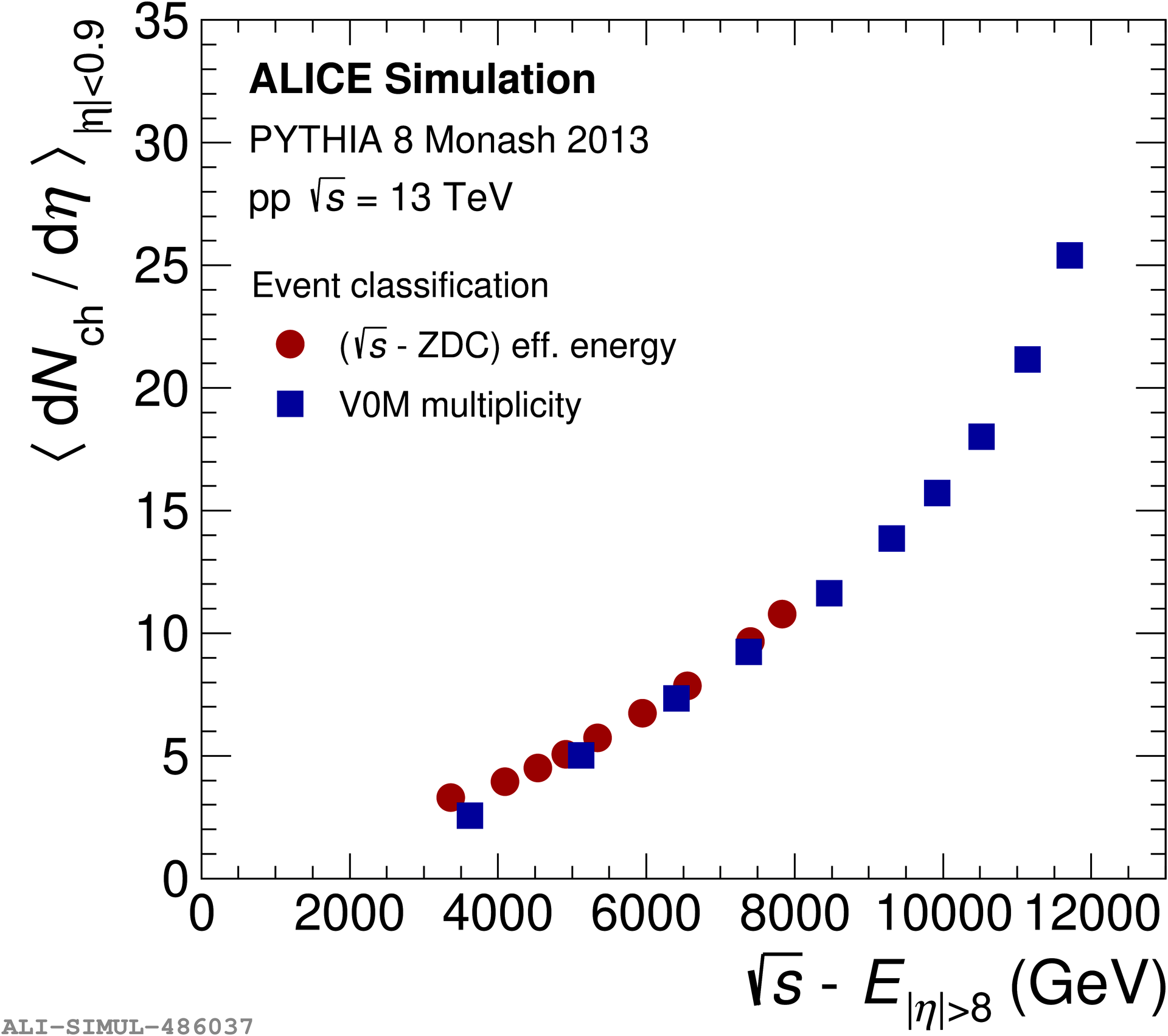}

		\vspace{14pt}
		\includegraphics[scale=0.105]{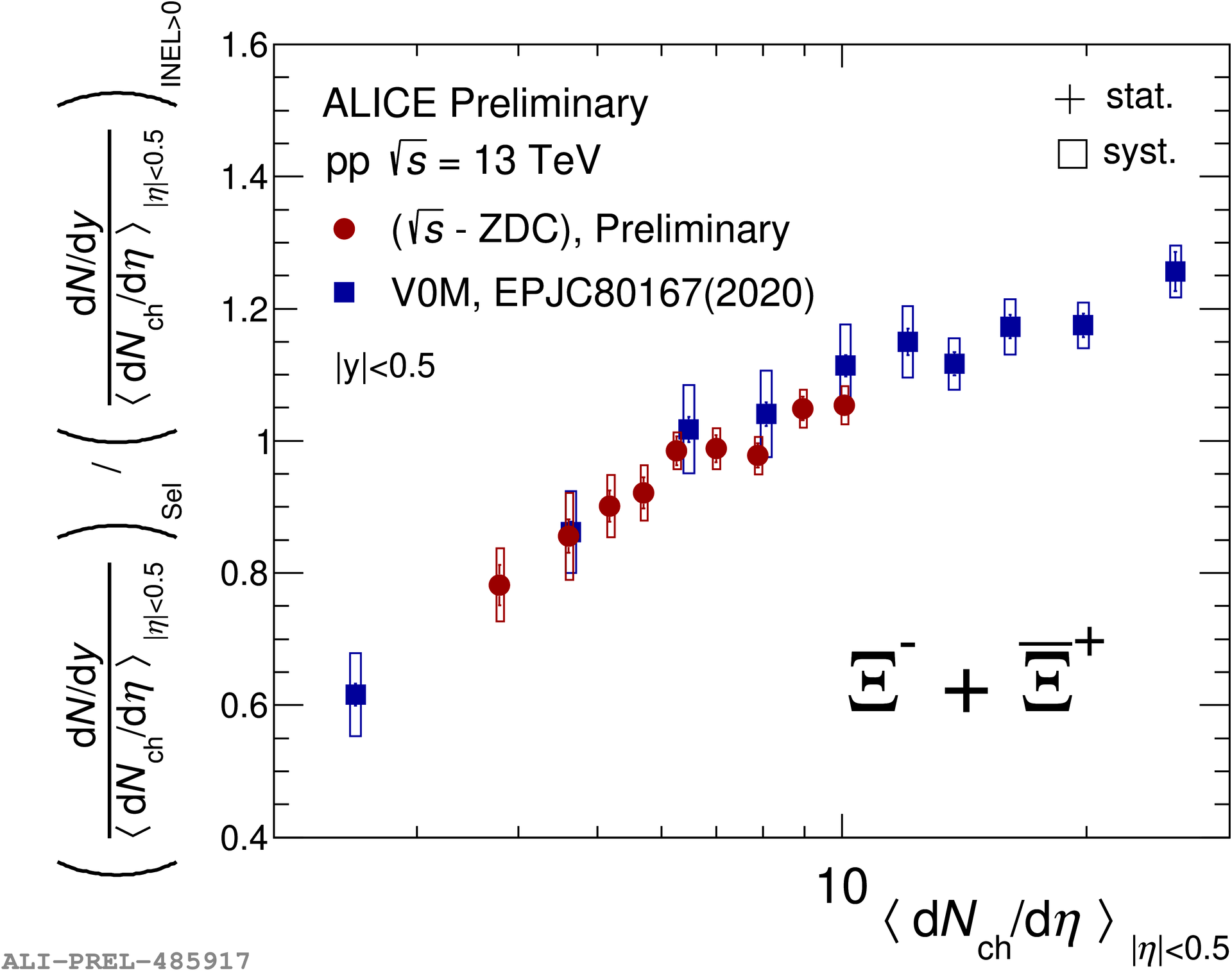}
		
		\flushleft	\texttt{\small {FIGURE 4. }}\small {Top: Multiplicity and effective energy correlation in PYTHIA 8 using V0 and ZDC event classes. Bottom: Ratio of $\mathrm \Xi$ yields to the charged-particle multiplicity (self-normalized to INEL$>$0) as a function of particle production at midrapidity, using V0 and ZDC standalone selections.}
	\end{center}

	\vspace{10pt}	

\hspace{-16pt} Fig. 4 (top plot) shows the simulation studies performed using Pythia 8 event generator and it has been observed that V0 and ZDC based event classes have sensitivity to multiplicity (final state) and initial effective energy. The self-normalized ratio of yields to the average charged-particle multiplicity (in INEL$>$0) with multiplicity selected through both V0 and ZDC are shown in Fig. 4 (bottom plot). The figure clearly shows that the strange particle production increases with multiplicity independent of the estimator used to classify the events. In other words, standalone analyses are not able to disentangle the initial and final state effects.\\

\hspace{-16pt} To distinguish the initial and final state contributions to strange particle production, V0 and ZDC combined classes are exploited using two approaches. In one approach, the events are selected using ZDC percentile selections, fixing the multiplicity through the V0 estimator in two classes (high multiplicity class and low multiplicity class) shown in Fig. 5 (right plot). In the other approach, the events are selected using V0 percentile selections while fixing the effective energy through the ZDC estimator to two classes (high effective energy class and low effective energy class) as shown in Fig. 5 (left plot). The figures suggest that there is no evolution with the effective energy and the final-state multiplicity is the dominant factor in strangeness enhancement.\\

\textbf{Conlusions:} \vspace{10pt}

\hspace{-16pt}The ALICE collaboration has performed studies for understanding strange particle production in high multiplicity pp collisions. The results suggest that strange hadron production has no significant evolution with initial effective energy and is driven by final particle multiplicity. Additionally, the study of strange hadrons in and out-of-jets suggests that soft processes are the dominant contribution to strange particle production. 
\end{multicols}
\vspace{10pt}
\begin{figure}
	\centering

	\includegraphics[scale=0.1]{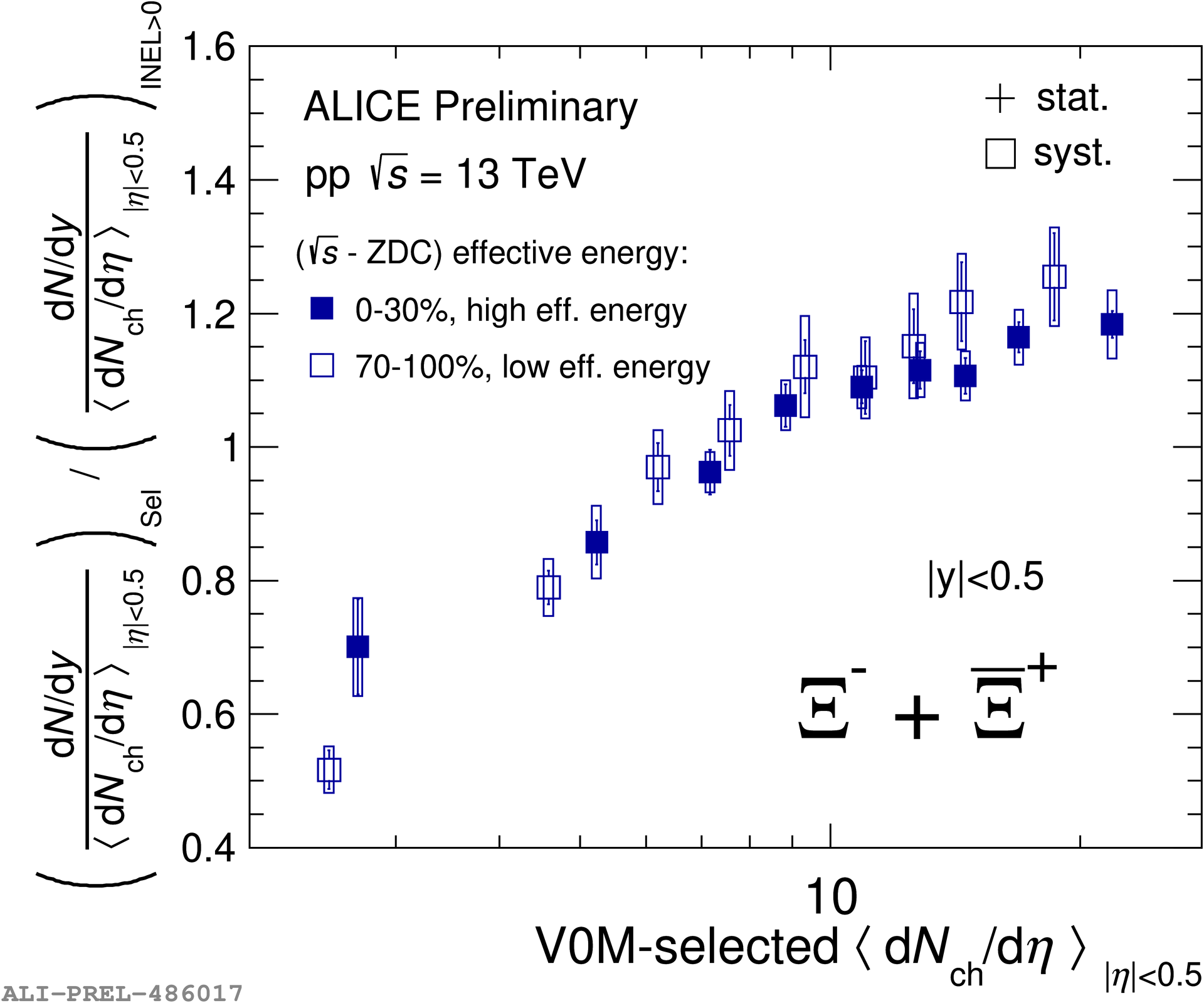}
	\hspace{0.5cm}
	\includegraphics[scale=0.102]{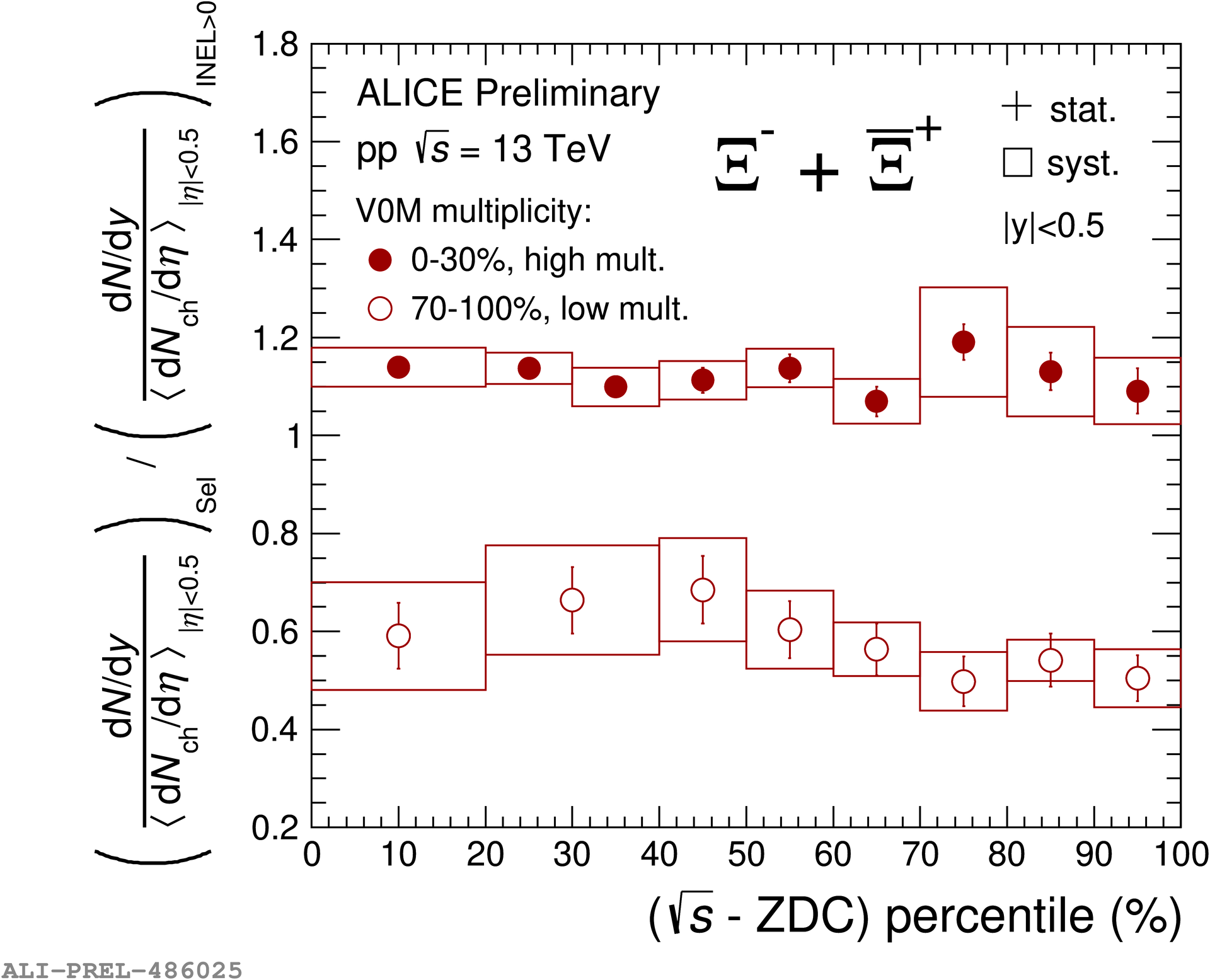}
	
\flushleft{	\texttt{\small {FIGURE 5.}}\small{ Ratio of $\Xi$ yields to the charged-particle multiplicity (self-normalized to INEL$>$0) extracted using V0-ZDC combined selections. Left: Events are selected using V0 event classes, fixing the multiplicity through the ZDC. Right: Events are selected using ZDC event classes, fixing the multiplicity through the V0 estimator.}  }
\end{figure}

\medline
\begin{multicols}{2}
	
\end{multicols}

\end{document}